\documentstyle[aps,preprint]{revtex}
\def\twiddle{\lower.9ex\rlap{$\kern-.1em\scriptstyle\sim$}}
\def\beq{\begin{equation}}
\def\eeq{\end{equation}}
\def\beqn{\begin{eqnarray}}
\def\eeqn{\end{eqnarray}}
\def\twiddle{\lower.9ex\rlap{$\kern-.1em\scriptstyle\sim$}}
\def\bigtwiddle{\lower1.ex\rlap{$\sim$}}
\def\gtwid{\mathrel{\raise.3ex\hbox{$>$\kern-.75em\lower1ex\hbox{
$\sim$}}}}
\title{Properties of Phase Transitions of a Higher Order}
\author{Pradeep Kumar\footnote{The address for 95--97:  Division
of Materials Research, National Science Foundation, 4201 Wilson
Boulevard, Arlington, VA  22230.}}
\address{Center for Ultra Low Temperature Physics, National High
Magnetic Field Laboratory and\\
Department of Physics\\
University of Florida\\
Gainesville, FL  32611  USA}

\begin{document}
\maketitle
\begin{abstract}
The following is a thermodynamic analysis of a III order (and some aspects 
of a IV order) phase transition.  Such a transition can occur in a
superconductor if the normal state is a diamagnet.  The equation for 
a phase boundary in an $H$--$T$ ($H$ is the magnetic field, $T$, the 
temperature) plane is derived.  by considering two possible forms of the
gradient energy, it is possible to construct a field theory which
describes a III or a IV order transition and permits a study of 
thermal fluctuations and inhomogeneous order parameters.
\end{abstract}
\section{Introduction}

If the Ehrenfest classification were used to describe a third or
a fourth order phase transition, the free energy and its low
order derivatives, e.g., entropy and the specific heat, will be
continuous, but, say for a III order transition\cite{pk1}, the
specific heat will have a discontinuous temperature derivative. 
For a IV order transition, the temperature derivative of the
specific heat will be continuous, rather, the second derivative of
the specific heat will be discontinuous.

In fact, this is not enough.  Consider a II order phase
transition\cite{pipp} boundary.  The phase boundary (in the
$H$--$T$ plane, where $H$ is the mechanical variable
 and $T$, the temperature) is given by
\beq
\left({dH\over dT}\right)^2 = {\Delta C\over T\Delta \chi}
\eeq
Here, $\Delta C = C_o - C_d$ where $C_o$ and $C_d$ are
respectively the ordered and disordered phase specific heat ($C =
-T {\partial^2 F\over \partial T^2}$) and $\Delta \chi =\chi_o
-\chi_d$ is the similar discontinuity in susceptibility ($\chi =
- {\partial^2F\over \partial H^2}$).  Having $\Delta C =0$ is,
naturally, not enough reason to argue for a higher order transition.  
A horizontal second order phase boundary may have $\Delta C =0$.  \ 
A higher order transition thus necessarily requires both $\Delta
C =\Delta \chi =0$.  \ Stated differently, near the transition
temperature, $F(T) = -(T_c -T)^n$ and $F(H) = - (H_c -H)^{n'}$.  The
Ehrenfest classification assumes that $n=n'$. In a scaling description
of a II order phase transition, $n = 2-\alpha$, where $\alpha$,
the specific heat exponent, is small.  It is, however, clear that
there is no limit on what $\alpha$ can be, whether positive or
negative.  Thus when $\alpha$ is negative and large in magnitude,
the transition is probably better described as with an order
corresponding to the nearest integer to $n$.  \ Any residue in
$n$ then can be viewed as a fluctuation contribution.

The well known liquid-vapor phase boundary for water is mostly 
I order, except when this phase line terminates at the critical 
point.  At the critical point, the transition is second order 
while past the critical point, the transition is continuous 
without any free energy singularities.  A possible scenario is that a
first order line may terminate at a higher order phase transition. 
We encounter such a possibility in the discussion of magnetic
field effects.

Recent observations in several high $T_c$ superconductors have
been cited as questioning whether the order of the transition is
2.  \ Probably the most extensive study\cite{good} comes for
Ba$_{0.6}$K$_{0.4}$BiO$_3$ (BKBO), a cubic superconductor with a
$T_c = 30$~K.  \ Here both specific heat\cite{hund} and
susceptibility\cite{good} have been measured and found to be
continuous at the transition.  There are
indications\cite{seid,juno} that specific heat in Bi and Tl based
cuprates has features characteristic of a higher order phase
transition.  The issue still needs to be resolved by careful
analysis of both the specific heat and the susceptibility.  In
particular, we need to know whether the transition is of order
III or IV.

In the following, we consider both a III order as well as a IV
order transition.  Section~II contains a thermodynamic analysis
of a higher order phase transition.  Section~III contains a field
theory, a Ginzburg Landau type free energy which describes such a
transition.  This section contains only properties associated
with a uniform order parameter.  A non-uniform order parameter is
the subject in Section~IV, whether caused by an external magnetic
field or thermal fluctuations.  Both are analyzed. 
Finally, Section~V contains a summary and some outlook of future.

\section{Thermodynamics}
The objective in this section is to derive an equation equivalent
to the Clausius-Clapeyron (C-C) equation for a higher order phase
boundary and discuss its
consequences.  There are three quantities, $\partial S\over
\partial T$, ${\partial S \over \partial H} = {\partial M\over
\partial T}$ and $\partial M\over \partial H$, each continuous
across the transition.  The usual derivation of the phase
boundary equation proceeds by equating the changes in each of the
above quantities as one moves along the phase  boundary.  The
resulting equalities are then solved for the slope of the phase
boundary.  Thus, the equations are (making frequent use of the
Maxwell relation ${\partial S\over \partial H} = {\partial M\over
\partial T}$):
\beq
{\partial H\over \partial T} = {\Delta \partial^2 S/\partial T^2
\over \Delta \partial^2 S/\partial H \partial T} = -{\Delta
\partial^2 M/\partial T^2 \over \Delta \partial^2 M/\partial T
\partial H} = {\Delta \partial^2 S/\partial H^2 \over \Delta
\partial^2 M/\partial H^2} = -{\Delta \partial^2 M/\partial T^2
\over \Delta \partial^2 S/\partial H^2}
\eeq
Of all of the possible expressions (some of which can be found in
Pippard\cite{pipp} the one below is special for two reasons: 
Firstly, the numerator contains thermal quantities and their
derivatives, while the denominator contains mechanical
expressions, similar to the C-C equation (Eq.~(1)) for a second
order transition.  Secondly, the equation has a symmetric form,
suggestive of a simple extension to higher order transitions.
\beq
\left[ {\partial H\over \partial T}\right ]^3 = -{\Delta
\partial^2 S/\partial T^2 \over \Delta \partial^2 M/\partial H^2}
= - {\Delta \partial C/\partial T \over T_c\Delta \partial
\chi/\partial H}
\eeq

Eq. (3) involves the first derivative of $\chi$ with respect to
$H$.  Thus the phase boundary here is determined by the nonlinear
susceptibility.  If the correction to the susceptibility is
quadratic in $H(\chi (H) =\chi_0 +\chi_2 H^2)$ then assuming that
specific heat field dependence is weak, the phase boundary can be
integrated to show that at low fields, $\Delta T_c \propto
H^{4\over 3}$, a result which is different from a conventional
superconductor where the Abrikosov result for $\Delta T_c$ is
linear in magnetic field.

Eq. (3) suggests the form of the phase boundary for a IV order
phase transition.  One might guess the (sometimes called
``Ehrenfest equation'') phase boundary in a IV order phase
transition to be
\beq 
\left( {\partial H\over\partial T}\right )^4 = {\Delta
\partial^2 C/\partial T^2 \over T_c \Delta \partial^2
\chi/\partial H^2}
\eeq

Going back to III order, given that we expect 
$F = -(T_c -T)^3$, the specific heat $C(T)$ is given by
\beq
C_{\hbox{III}} (T) = -T \partial^2F/\partial T^2 \simeq T(T_c-T)
\eeq
The specific heat in fact has a broad ``peak'' at $T_c/2$.  \ A
transport measurement might report a true $T_c$ while a
thermodynamic measurement will report a smaller $T_c$.  \
Moreover, if, as is customary, the calorimetric measurement of
$T_c$ is made by deriving $T_c$ from a point where the entropies
(of the order and disordered state) match, one will have yet
another value for $T_c$.  \ The essential fact is that the
transition is not broad, it is described by a discontinuity in
higher order thermodynamic derivatives.

Looking back at Eq. (1), we note another curious thermodynamic
fact.  Since a typical metal is paramagnetic and a superconductor
diamagnetic, it is not possible to have $\Delta \chi =0$. \ In
fact this condition is satisfied only when the normal state is a
diamagnet, as is the case with all of the material examples
mentioned above.  {\it A higher order transitions is possible only
if the normal state is a diamagnet.}

Finally, we can estimate the thermodynamic critical fields. 
Thus, in the standard way
\beq
{B_c^2\over 2\mu_0} = (T_c-T)^n
\eeq
i.e., $B_c \sim (T_c-T)^{n/2}$, a result that is reasonable
in the well known II order case but entirely unexpected for the
higher order phase transitions.  In the following we construct
specific models to better understand the possible microscopic
origins of these results.

\section{Ginzburg-Landau Theory}
For a III order phase transition, a Ginzburg-Landau theory
contains some surprises even though it is relatively
straightforward to obtain the characteristic form of the free
energy $F_h(\Delta, T)$, here $\Delta$ is the order
parameter. $F_h$ refers to the free energy with a homogeneous
order parameter and $F_0$ is the normal state free energy.
\beq
F_n = F_0 + a\Delta^4 + b\Delta^6
\eeq
Here $a = a_0 ({T\over T_c} -1)$ and $b\geq 0$ a constant. 
Searching for the minimum of the free energy as a function of
$\Delta$, we find
\beqn
\Delta^2_0 = {2|a|\over 3b} = {2 a_0 \over 3b} (1 - {T\over
T_c})&~& \quad T < T_c\nonumber\\
= 0 &~&\quad T > T_c
\eeqn
The various thermodynamic quantities\cite{ttc} (all of the 
quantities refer only to the condensing degree of freedom) are given by
\beqn
\langle F\rangle & = & {4\over 27} \cdot {|a|^3\over b^2} =
{4\over 27} \cdot {a_0^2 \over b^2} \left(1 - {T\over T_c}\right
)^3\\
S & = & {\partial \langle F\rangle\over \partial T} = -{4\over 9}
\cdot {a_0^3\over b^2T_c} \left ( 1 - {T\over T_c}\right )^2\\
C & = & {\partial^2 \langle F\rangle\over \partial T^2} = {8\over
9} \cdot {a_0^3\over b^2T_c^2} T\left (1 - {T\over T_c}\right )
\eeqn
We see that the specific heat has a characteristic temperature
dependence.  It shouldn't be taken for a broad discontinuity 
of a II order transition
since ${\Delta T\over T_c} \sim {1\over 2}$; where $\Delta T$ is
some measure of the transition width.

We should also recall now that there is no quadratic term in the
order parameter in Eq.~(7).  In the presence of a quadratic term,
the transition becomes a I or a II order\cite{dubo}.  This is
a rather important and subtle point.  For example, the magnetic
transition in solid $^3$He\cite{dub2} involves interaction
energies that do not have a quadratic term, they have only a
quartic term.  However, there is always a quadratic contribution
from the entropy.  Thus the transition is usually a I order one.  
A third order transition requires vanishing quadratic term in the 
free energy.  One possible
interpretation of Eq. (7) can be that the density of states at
the Fermi surface (if Eq.~(7) is the free energy for a
superconductor) which appears as an energy scale determining
factor, is in fact proportional to $\Delta^2$.  \ Thus, we could
imagine a transition between an insulator and a superconductor. 
The Fermi surface density of states is zero in the normal
state indicating an insulator.  It
is finite in a metal/superconductor and the transition would be a
curious but a profound feedback phenomena.  As far as I am aware,
there are several quadratic terms in a real
insulator-superconductor transition and the real
transition is a robust II order one.  The above analogy is only
an illustration.

What happens if the transition is a IV order?  A possible free
energy can be written\cite{sasl} as
\beqn
F_h & = & F_0 + a|\Delta|^6 + b|\Delta |^8\\
a & = & a_0 (T/T_c -1)\nonumber
\eeqn

Here, one might imagine that the prefactor density of states at
the Fermi surface is driven by superconductivity to be quartic in
the order parameter.  The results corresponding to Eqs~(8--11)
are
\beqn
\Delta_0^2 & = & + {3a_0\over 4b} (1-T/T_c) \qquad a < 0\\
\langle F\rangle & = & -{27\over 256} {|a|^4\over b^3}\\
C_{\hbox{IV}} (T) & = & {27\over 64} {T\over b^2 T_c^2} (1 -
T/T_c)^2
\eeqn

\section{Spatially Inhomogeneous Order Parameter}
Here we consider two effects, first the effect of magnetic field
and then the effect of thermal fluctuations.  To do either, we
need to supplement the free energy Eq.~(7) with terms that
involve a spatially varying order parameter.  Again there are many
choices and it is less obvious how they could be eliminated.  The
following contains consequences of one selection, we will comment
on another selection, briefly, at the end of this section.

\subsection{Model I}
Thus, we supplement the free energy (Eq.~(7)) by
\beq
F_G = C|\nabla\Delta|^2
\eeq

To be specific, we will assume that $\Delta (\twiddle r)$ is a
complex function of position in 3d.  Let us first look at the
temperature dependent correlation length.  The Euler-Lagrange
equation for the order parameter is given by
\beq
{\delta \over \delta\Delta} (F_h + F_G) = -2|a| \Delta^3 + 3b
\Delta^5
- C \nabla^2\Delta = 0
\eeq
We look for small amplitude variations of $\Delta(r)$ by writing
\beq
\Delta (\twiddle r) = \Delta_0 + \delta (\twiddle r)
\eeq
The linear equation determining $\delta(r)$, the final solution
being dependent on boundary conditions, is
\beq
-C \nabla^2 \delta (r) + \left\{ {1\over 2} \left.{\delta^2
F_h\over \delta\Delta^2}\right|_{\Delta_0} \right\} \delta (r) =0
\eeq
leading to a length scale, the temperature dependent correlation
length $\xi(T)$
\beq
\xi^2(T) = C \left[ {1\over 2} \left. {\delta^2 F_h\over \delta
\Delta^2}\right |_{\Delta_0}\right ]^{-1} = 3 bC/8|a|^2 
\eeq

Thus, in contrast to the familiar notion of length scale, here
there is an asymmetry about $T_c$.  \ For $T>T_c$, the
correlations are non-exponential.  Below $T_c$, the correlations
length depends linearly on $1/|a|$, as opposed to the square root
dependence in a second order phase transition.

We can also identify the superfluid density.  A current can be
viewed as that arising from the space gradients.  In that case,
the density is given by expressing the free energy due to current
as $V_2\rho_sv_s^2$ where $v_s$ is the supercurrent.  With
Eq.~(16), we have $\rho_s \sim \Delta^2$ and the consequent
temperature dependence.
\vspace{.15in}

\begin{centering}
{\it A.1.\ \ Magnetic Field Dependence}
\end{centering}

The gauge invariant free energy requires that the charge
interaction with magnetic field be described by a transformation
of the gradient term
$\nabla\to (\nabla + {2 \pi i \over \phi_0}\twiddle A)$ here $\phi_0$
is the flux quantum $h/2e$ and $A$ is the vector potential($B =
\nabla \times A$ refers to the local field, which approaches the 
external field H outside).

It is straightforward to show that the equation for the vector
potential represents flux expulsion whenever $\Delta_0\neq 0$. 
It is also true that in analogy with a conventional
superconductor, $\lambda^{-2} \propto \Delta^2$.  \ All these
features are standard and are dependent on $n_s = \Delta^2$.  \
The difference appears, not entirely unexpectedly, at the
calculation of $H_{c2} (T)$.  \ The curious result is that the
transition becomes I order in the presence of a magnetic field.

To see that, we derive the criterion for the instability of the
normal state.  The lowest order Euler-Lagrange equation is given
simply by
\beq
-C \left| \left ( \nabla +{2\pi i\over \phi_0} A\right )\right|^2
\Delta = \lambda \Delta
\eeq

The solutions of this equation are dependent on the choice of
gauge, but the eigenvalues $\lambda = (2n +1) {2\pi C\over
\phi_0} H$ is independent of gauge choice, with the lowest energy 
level corresponding to $n=0$.  
If the corresponding wave function is $\phi_n (\twiddle r)$, we 
can write
\beq
\Delta(\twiddle r) =\sum_{(n)} \zeta_n \phi_n (\twiddle r)
\eeq
where $n$ stands for the set of quantum numbers necessary to
characterize the state.  In the mean field sense, the free energy
can be written as
\beq
F = {2\pi C\over \phi_0} H \zeta_0^2 + a \zeta_0^4 + b \zeta_0^6
\eeq
The transition occurs at $H_{c2}(T)$ where
\beq
H_{c2}(T) = {\phi_0\over 8\pi bC} |a|^2
\eeq

However, the order parameter develops discontinuously at $T_c$. 
Its value at $T_c$, $\Delta \zeta_0$ is given by
\beq
\Delta \zeta_0 = \left ( {2\pi C\over b\phi_0} H\right )^{1\over
4}
\eeq
The latent heat ($L=T_c \Delta s$) can also be obtained, it is
given by
\beq
L = T_c \Delta s = {2\pi a_0 C\over b\phi_0} H/T_c
\eeq
All of these quantities are measurable and provide an unequivocal
test of the model for the gradient energy.  They represent a
first order phase transition for $H\neq 0$ while the $H=0$
transition is a III order one.  We have a first order line ending
in a critical point which is of higher order than the
conventional second order.

At this stage, the applicability of ideas of this subsection
to BKBO breaks down.  As far as we can tell, the transition is
{\it not\/} of I order in any finite field, there is no latent
heat, no discontinuity---nor any other trace.  In fact, all
evidence in finite field, points to a transition higher in order
than II.  \ Before proceeding with another ansatz for free energy
in sec.~4.2, let us consider thermal fluctuations following Eq.
(16).
\vspace{.15in}

\begin{centering}
{\it A.2.\ \ Thermal Fluctuations}
\end{centering}

Let us consider a scalar order parameter $\Delta (\twiddle r)$.  We
consider a partition function
\beq
z =\int D[\Delta (\twiddle r)] \exp [-\beta (F_h + F_G)]
\eeq
The partition function provides the thermodynamic free energy
which then leads to the other thermodynamic properties.  As is
customary, the integral is evaluated in a saddle point
approximation.  Consider $\Delta (\twiddle r) = 
\Delta_0 + \delta (\twiddle r)$
where $\Delta_0$ is the minimum of $F_h$, as in Eq.~(8):
\beq
e^{-\beta \hat F} = z = e^{-\beta\langle F_h\rangle} \prod_k \int
d \delta_k \exp\left[-{\beta\over 2} \sum_q\left\{\left. {\delta^2
F_h\over \delta\Delta^2}\right|_{\Delta_0} + cq^2\right \}
\delta_q^2\right]
\eeq
where $\delta_q$ (or $\delta_k$) is the Fourier transform or
$\delta(\twiddle r)$, and the first term is given by $8|a|^2/3b$.  
\ Thus, the free energy becomes
\beqn
\hat F & = & \langle F_h\rangle + F_f\nonumber\\
F_f & = & -{kT\over 2} \sum_q ln \left [ 2\pi kT \bigg /
\left({8|a|^2\over 3b} +Cq^2\right )\right ]\\
C_f & = & -T{\delta^2 F\over \delta T^2} \cong |a|^{d-2}
\eeqn
In this model, the transition at $H=0$ is quite robust.  In
Eq.~(30), we have the temperature dependence of the fluctuation
specific heat, which is well behaved for $d\geq 2$.  \ The upper
critical dimension for this model is $d=2$.  \ (Recall that
the upper critical dimension in a II order case is $d=4$).  \ The
fluctuations in this model, described by Eq.~(16) are divergent
only for $d=1$.  \ Thus there are no critical fluctuations and no
need to go beyond mean field theory, at least not for $d\geq 2$.
This also means that if the experimental exponents are not 
meanfield like, model I is probably not applicable. 

If $\Delta (\twiddle r)$ were a 2--d vector, then the phase
fluctuations behave similar to their behavior in case of a II
order phase transition.  There are the usual infrared divergences
in 1 and 2--d indicating the importance of topological 
defects\cite{kost} in the phase transitions, etc.

\subsection{Model II}
Let us consider a different gradient energy term.  If we follow
the physical picture where the overall energy scale in the free
energy depends on the order parameter, the gradient term perhaps
looks like
\beq
F_G = \tilde c \Delta^2 |\nabla\Delta|^2
\eeq
which can also be viewed as the square of the gradient of
$\Delta^2$.  \ We see that for a scalar order parameter, the
Eyler-Lagrange equation is given by
\beq
-2|a|\Delta^2 + 3b \Delta^4 -\tilde c(\nabla\Delta)^2 - \tilde c
\Delta \nabla^2 \Delta =0
\eeq
A linearized version of this equation, about the homogeneous
solution, becomes
\beq
4|a| \delta (r) -\tilde c \nabla^2 \delta (\twiddle r) = 0
\eeq
In contrast to Eq. (19), the correlations length exponent is back
to being $1\over 2$.  \ But much else has changed.  The effect of
magnetic field can be incorporated by replacing the derivative in
Eq. (31) by a covariant term.  Like sec.~IV.A.1, the order
parameter can be expanded in the Landau orbital wave functions
which are eigen functions of the covariant Laplacian.  The
nonquadratic nature of the free energy introduces interactions
and transitions between the Landau orbital states.  However
overlooking these interactions (for small fields as in
sec.~IV.A.1), we see that the transition in the presence of a
magnetic field is III order with a phase boundary given by
\beq
{2\pi\tilde c\over \phi_0} H= a_0 (1 - T/T_c)
\eeq
If $\Delta (r)$ is a complex number, the number density of
excitations is proportional to $\Delta_0^4$, a feature also
shared by the penetration depth ($\lambda^{-2} \propto
\Delta^4$).

The proverbial fly in this ointment comes from a study of thermal
fluctuations.  The thermal fluctuations contain a divergent
(logarithmic) contribution to the free energy, yielding a
specific heat that diverges as $|a|^{-2}$, independent of 
dimension.  There is additional
dimension dependent divergence, which for specific heat takes the
form $C \propto |a|^{(d/2-2)}$.  To see the details, consider
Eq.~(28) as worked out for model II, we have
\beqn
e^{-\beta\hat F} &=& z = e^{-\beta\langle F_G\rangle} \prod_k
\int d \delta_k \exp \left [ -\beta \sum_q \Delta_0^2 [4|a| +
cq^2] \delta_q^2 \right]\\
\hbox{i.e.}\qquad F_f &=& {kT\over 2} \sum_q ln \left[ \Delta_0^2
( 4|a| + cq^2)\right ]
\eeqn
It is the $\Delta_0^2$ term that is singular in the eventual
derivation of the entropy and specific heat.

At this moment, it is hard to imagine that the entire mean field
analysis of model II, as described above, is meaningless.  And
yet, a mean field analysis depends on the validity of the saddle
point evaluation of the partition function.  An estimate of
corrections to mean field, seen above for model II, is divergent
and it is unclear if some sort of renormalization will restore
these results to a finite value.  Barring that, model II remains
very dubious as a valid model.  These problems are equally
exacerbated for a IV order phase transition.

\section{Summary}
It appears therefore that a thermodynamic description of a III or
IV order phase transition is relatively straightforward, although
the consequences are almost unavoidably subtle.  Once we
recognize the possibility of these phase transitions, it is clear
that we cannot lump them together with all of the other continuous
transitions.  There are subtle and interesting differences which
call for a separate identification and analysis.

It is possible that a higher order transition has been observed
before.  Then it was attributed to an extreme case of sample
inhomogeneity since the transition temperature measured from
different techniques were all different.  The above is an attempt
to outline a systematic formalism for the analysis of higher
order transitions.  We have an equation for the phase boundary
Eqs.~(3) and (4) and furthermore, we also have a field theory
which leads to a III (or IV) order phase transition.  In
searching for the effects of a magnetic field or thermal
fluctuations, we find two models.  In one case, the finite field
transition is a I order transition which ends into a III order
critical point.  All thermodynamic quantities vanish at this
critical point ($H=0$) appropriately.  In the other model, we
find strong thermal fluctuations which, on the one hand appear to
question the validity of a mean field theory.  It is also
possible that the correct model still remains elusive.

\section*{Acknowledgments}
This paper was partially supported by DOEDE--FG05--91ER45462.  

I am grateful to several for discussions, particularly, R.~Goodrich, 
V.D. Mineev, W.~Saslow and H.~Suhl.


\begin{references}
\bibitem{ttc} Strictly speaking, the entropy should have a
constant subtracted so that $S( T=0) =0$.  This would yield a net
value of the entropy at $T = T_c$ which can then be used in
determining the transition temperature, in case the specific heat
discontinuity is unusually broad.
\end{references}
\end{document}